\documentclass[letter]{aa} 
\usepackage{graphicx}
\usepackage{txfonts}
\usepackage{natbib}
\begin{document}
\title{SOFIA observations of CO~(12--11) emission along the \object{L1157}
  bipolar outflow}  

\author{Jochen Eisl\"offel  \inst{1}
   \and Brunella Nisini     \inst{2}
   \and Rolf G\"usten       \inst{3}
   \and Helmut Wiesemeyer   \inst{3}
   \and Antoine Gusdorf     \inst{3,4}
       }

\institute{Th\"uringer Landessternwarte, Sternwarte 5, D-07778 Tautenburg,
  Germany, \email{jochen@tls-tautenburg.de}
\and
   INAF-Osservatorio Astronomico di Roma, Via di Frascati 33, I-00040,
   Monteporzio Catone, Italy
\and
     Max-Planck Institut f\"ur Radioastronomie, Auf dem H\"ugel 69, D-53121
     Bonn, Germany
\and
     LERMA, UMR 8112 du CNRS, Observatoire de Paris, \'Ecole Normale 
     Sup\'erieure, 24, rue Lhomond, F-75231 Paris Cedex 05, France}

   \date{Received end of January 2012; accepted ???}

 
  \abstract
   {Carbon monoxide is an excellent tracer of the physical conditions of gas
     in molecular outflows from young stars.}
   {To understand the outflow mechanism we need to investigate the
     origin of the molecular emission and the structure and interaction of the
     outflowing molecular gas. Deriving the physical parameters of the gas
     will help us to trace and understand the various gas components in the
     flow. }
   {We observed CO~(12--11) line emission at various positions along the L1157
     bipolar outflow with GREAT aboard SOFIA.}
   {Comparing these new data with CO~(2--1), we find basically constant line
     ratios along the outflow and even at the position of the source. These
     line 
     ratios lead us to estimates of 10$^5$ to 10$^6$\,cm$^{-3}$ for the gas
     density and 60 to 100\,K for the gas temperature of the outflowing gas.}
   {The constrained density and temperature values indicate that we are mostly
     tracing a low-velocity gas component everywhere along the outflow, which
     is intermediate between the already known cold gas component, which
     gets
     entrained into the flow, and the hot gas, which gets shocked in the 
     outflow.}


  \maketitle


\section{Introduction}

Carbon monoxide is widely considered to be the best probe of excitation 
conditions in the interstellar and protostellar molecular gas, because of 
its high abundance and simple chemistry. In the study of outflows from
young protostars, in particular, the CO molecule plays a central role. 
Low-$J$ rotational CO lines are routinely used to study the kinematics and
energetics of the molecular gas swept-out by the protostellar jets and
winds. At the same time, high-$J$ rotational CO lines, observable in the
far-IR wavelength range, are ideal 
tracers of the gas that is currently being shocked in the interaction
of the jets with the ambient medium, which significantly contributes 
to the overall shocked gas cooling (e.g. \citealt{giannini2001}).

In this contribution we report the first observation of a higher-$J$ line 
of CO, namely the CO~(12--11) line, along various shocked spots in the 
\object{L1157} bipolar outflow that were obtained with the GREAT heterodyne
spectrometer\footnote{GREAT is a development by the MPI f\"ur Radioastronomie 
and the KOSMA / Universit\"at zu K\"oln, in cooperation with the MPI f\"ur
Sonnensystemforschung and the DLR Institut f\"ur Planetenforschung.} 
(\citealt{heyminck2011}) aboard the Stratospheric Observatory for Infrared
Astronomy, SOFIA, during three flights in its Early Science phase
(\citealt{young2011}). 

The CO~(12--11) transition lies at 365\,K above the ground state and it is
therefore a very good tracer of warm molecular gas. As such, it can be used to
bridge the gap between the low-lying CO lines, which are related to the
swept-out outflow material, and the higher-$J$ lines excited in the shocked
hot gas. 

The \object{L1157} outflow is an ideal target for this pilot study with SOFIA
because it 
is one of the most active and extensively studied outflows from low-mass young
protostars. Observations in a wide range of wavelengths indicate the presence
of strong chemical and physical gradients within the flow. Chemically, the
outflow is extremely rich, with more than 20 molecular species
detected, showing significant morphological variations along the flow
(e.g. \citealt{bachiller2001}). This chemical differentiation is associated 
with the presence of shock spots that show a range of excitation
conditions that are probed, e.g., by multiwavelength H$_2$ observations
(\citealt{daviseisloeffel1995}; \citealt{caratti2006}; 
\citealt{nisini2010a}). These shocked regions show up prominently in 
H$_2$O emission as well, indicating that a warm gas-phase chemistry 
is actively taking place (\citealt{nisini2010b}).

The outflow has been fully mapped in CO~(2--1) by \citet{bachiller2001}, while
sub-mm CO observations have so far only been obtained in the 
blueshifted southern lobe. In particular, \citet{hirano2001} have obtained 
spectra of CO~(6--5), CO~(4--3), and CO~(3--2) in three different shock
positions, suggesting that the bulk of the emission comes from gas at $n \sim$
10$^4$\,cm$^{-3}$ and $T \sim$ 50 -- 150\,K. HERSCHEL observations obtained by 
\citet{codella2010} and \citet{lefloch2010} suggested that the CO~(5--4) 
emission in the brightest blueshifted shock spot (called B1 after
\citet{bachiller2001})  
can be divided into different velocity components with different excitation 
conditions: the low-velocity gas at $n \sim$ 3$\times$10$^5$\,cm$^{-3}$, 
$T \sim$ 100\,K, and the high-velocity gas at $n \sim$ 10$^4$\,cm$^{-3}$ 
and $T >$ 400\,K.

CO transitions at higher J (from J = 15 up to J = 20), which probe the hottest
gas  
component, have been detected by ISO at low spatial resolution and suggested
the presence of CO gas with temperatures in the range 300 -- 1000\,K
(\citealt{giannini2001}). Very recently, high spatial resolution, non-velocity
resolved PACS observations of CO lines from J = 14 to 22 have been reported 
for the \object{L1157-B1} position, which allows one to localize the hot gas in 
a compact shock of about 9$''$ in size and to better constrain its physical
properties (\citealt{benedettini2012}). 


\begin{figure}
\centering 
\includegraphics[width=11.5cm,angle=-90]{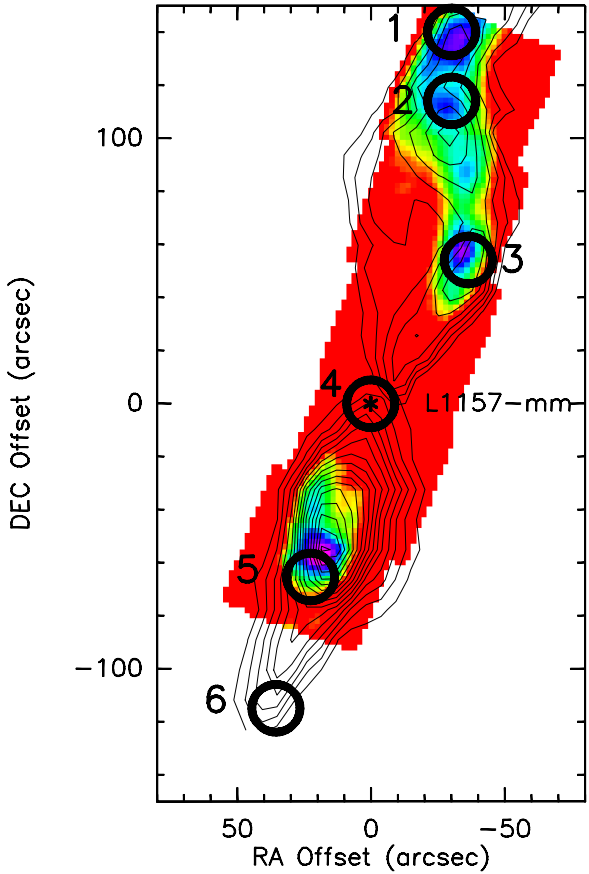}
\caption{Positions observed with SOFIA in the \object{L1157} bipolar outflow,
  which is emanating from the embedded Class\,0 source \object{L1157-mm}. The
  background colour 
  map shows an image in the H$_2$ 0--0~S(1) line at 17$\mu m$ (from
  \citealt{nisini2010b}) with superposed contours of the integrated CO (2-1)
  emission (from \citealt{bachiller2001}). The targeted positions of the
  presented CO~(12--11) observations with GREAT are labelled and indicated 
  with circles, with diameters equal to the actual beam size of 21\farcs3 of 
  the GREAT beam. 
\label{positions}}
\end{figure}


\begin{figure}
\centering 
\includegraphics[width=4.45cm]{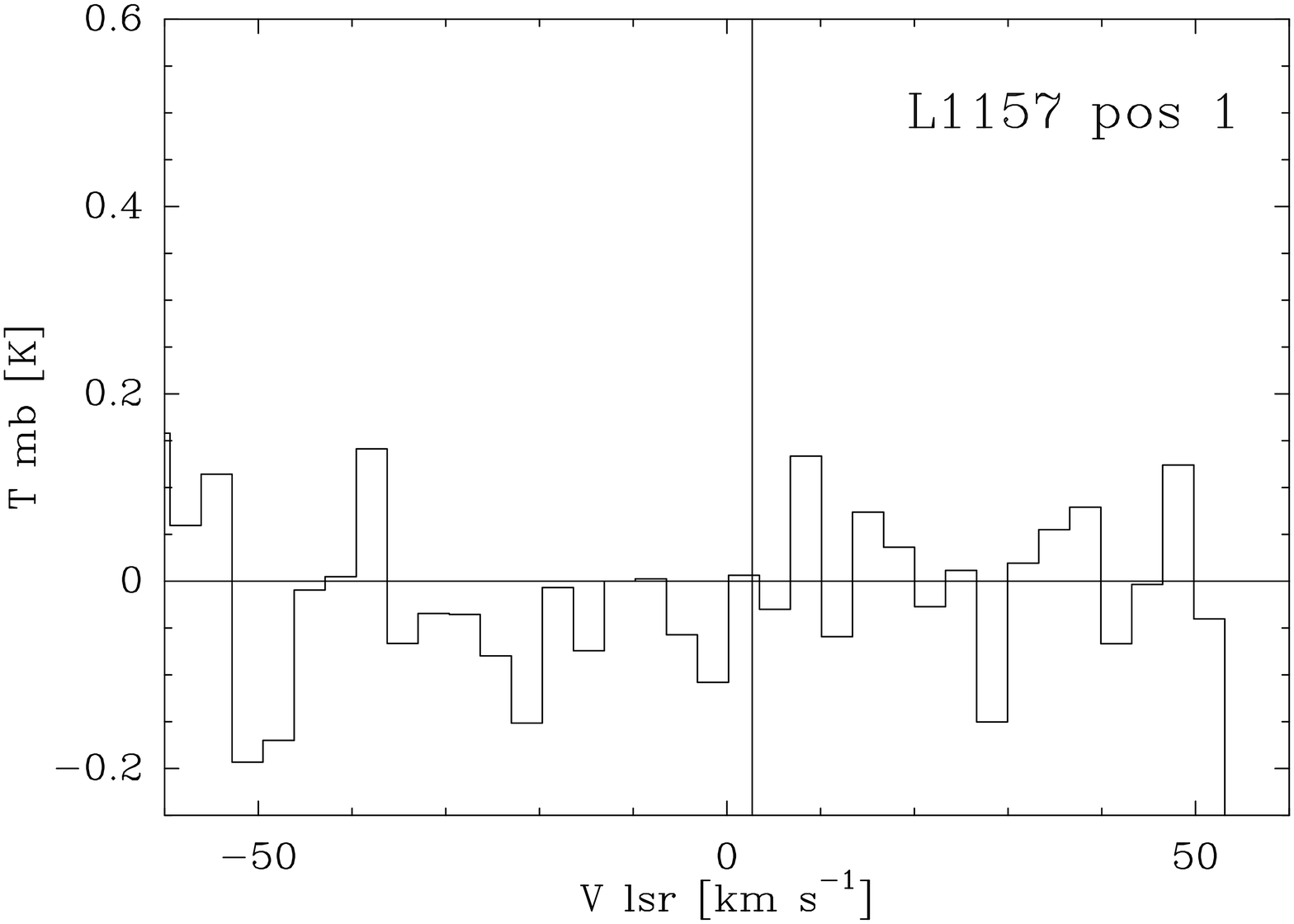}
\includegraphics[width=4.45cm]{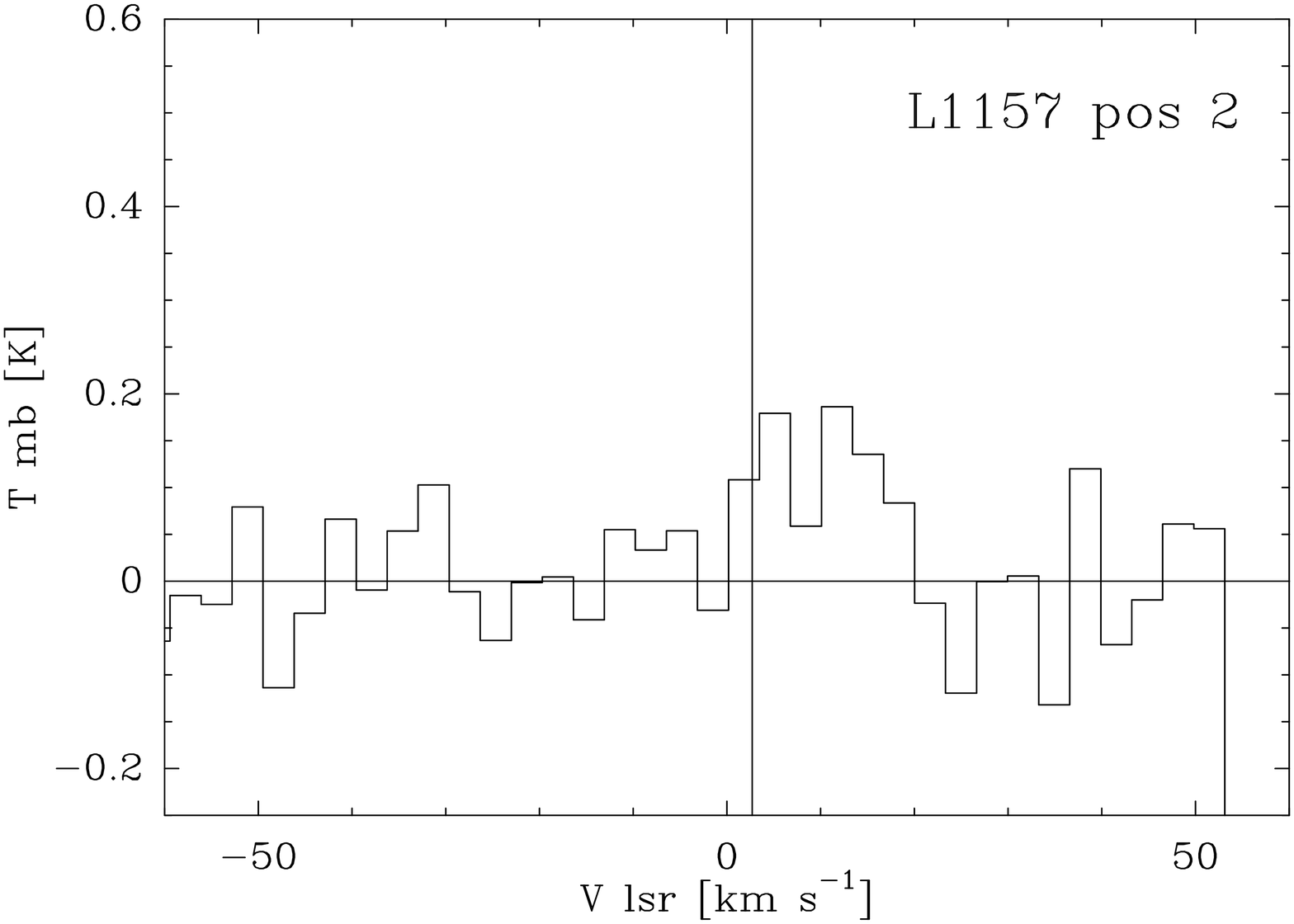}
\includegraphics[width=4.45cm]{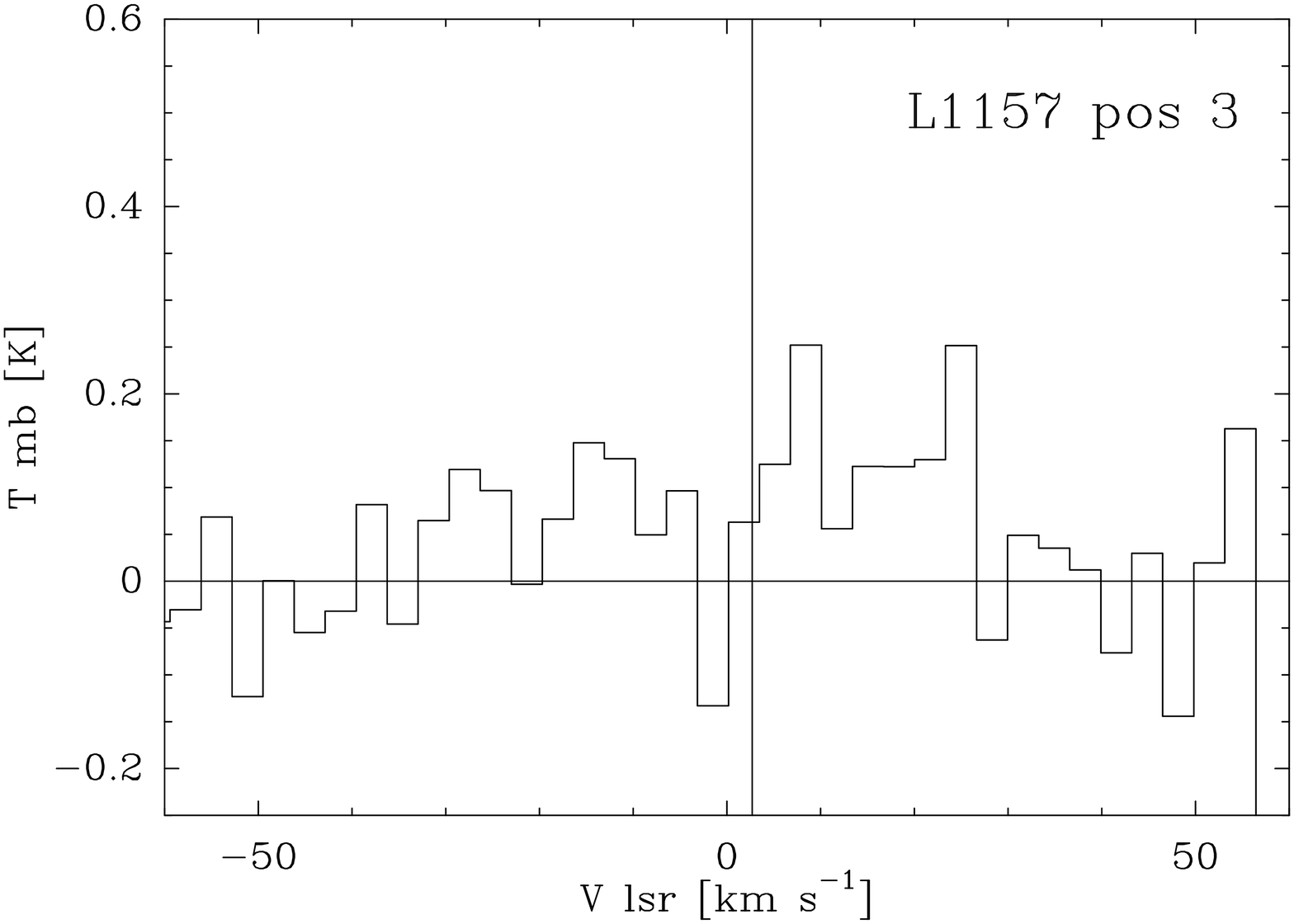}
\includegraphics[width=4.45cm]{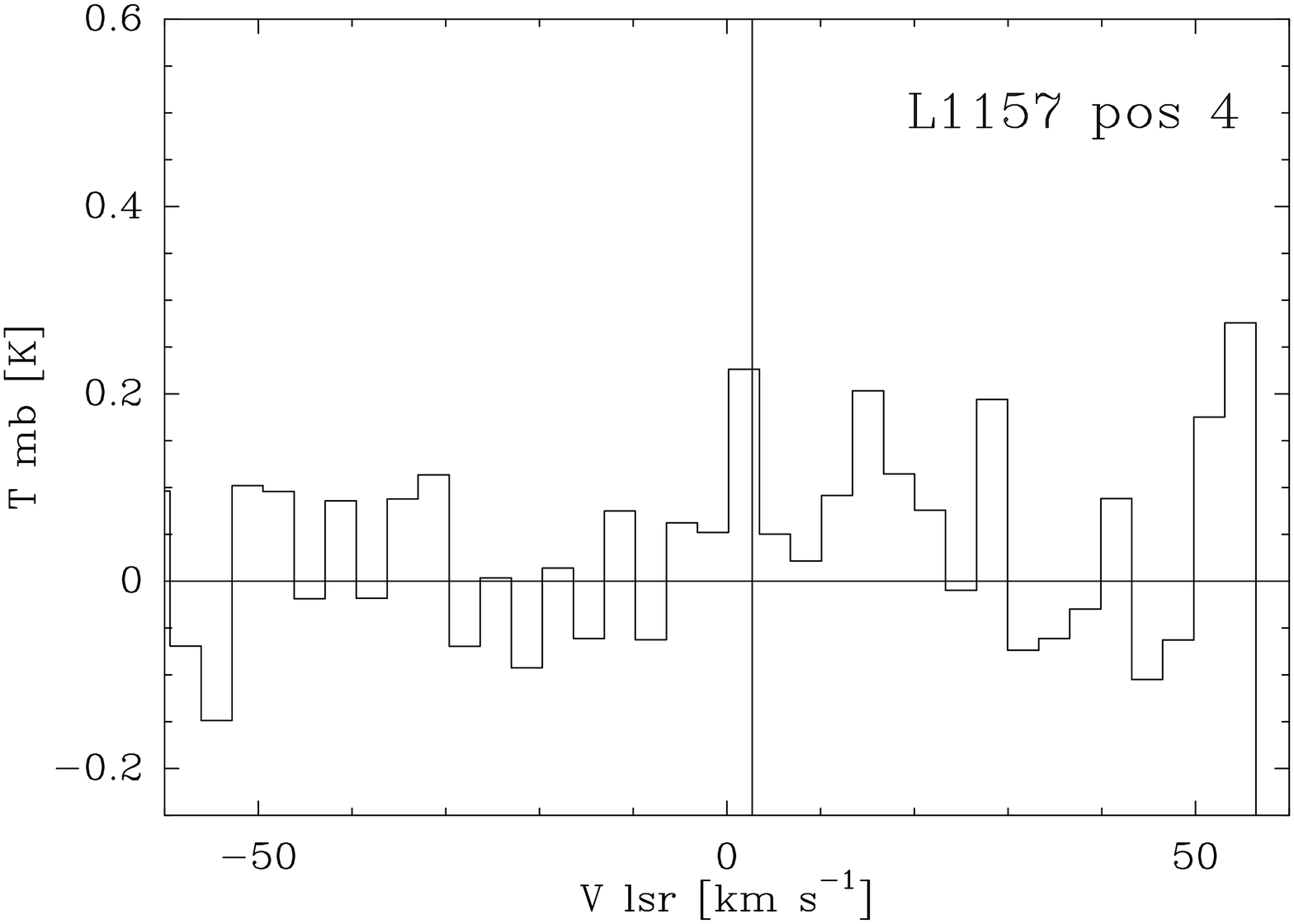}
\includegraphics[width=4.45cm]{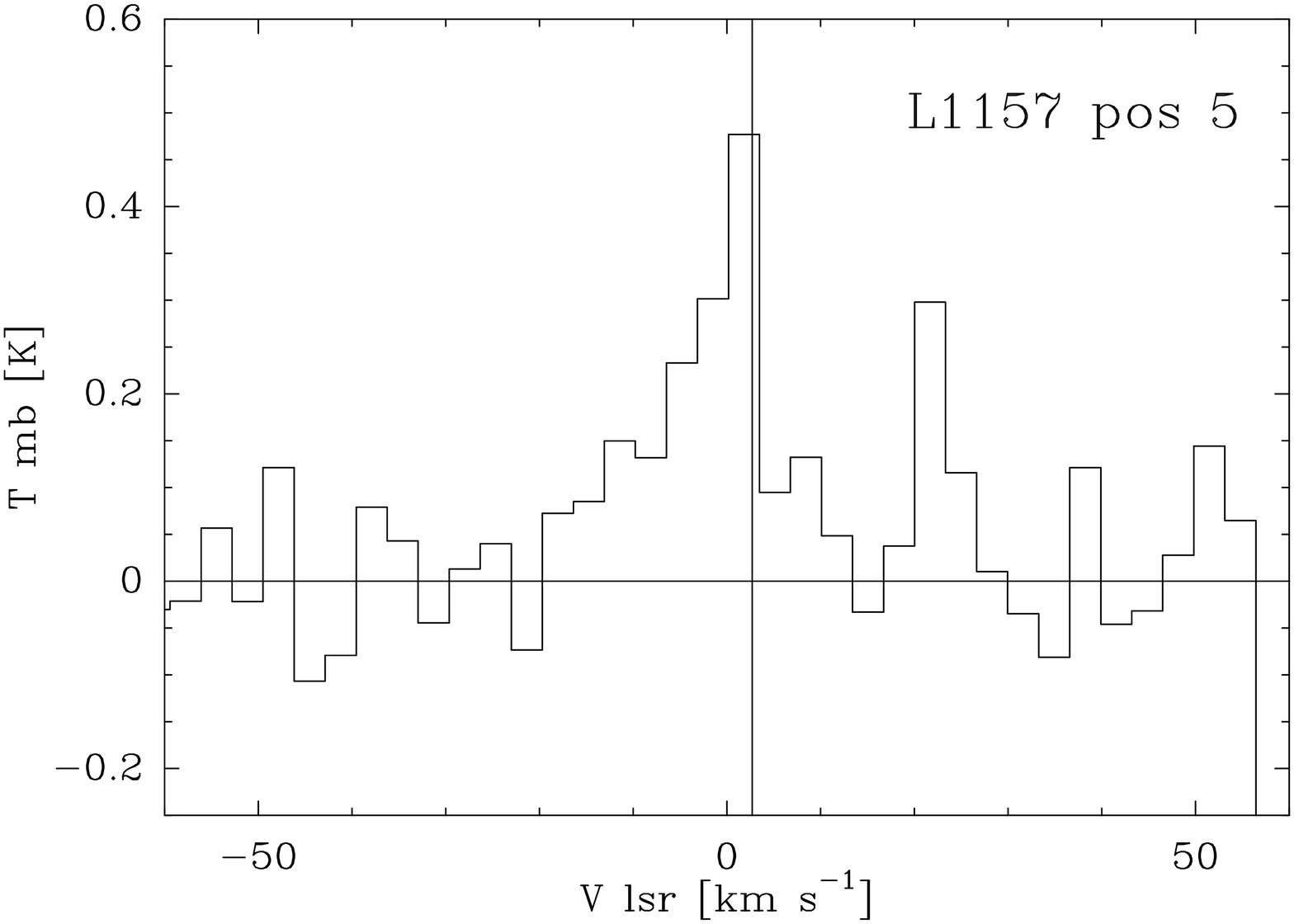}
\includegraphics[width=4.45cm]{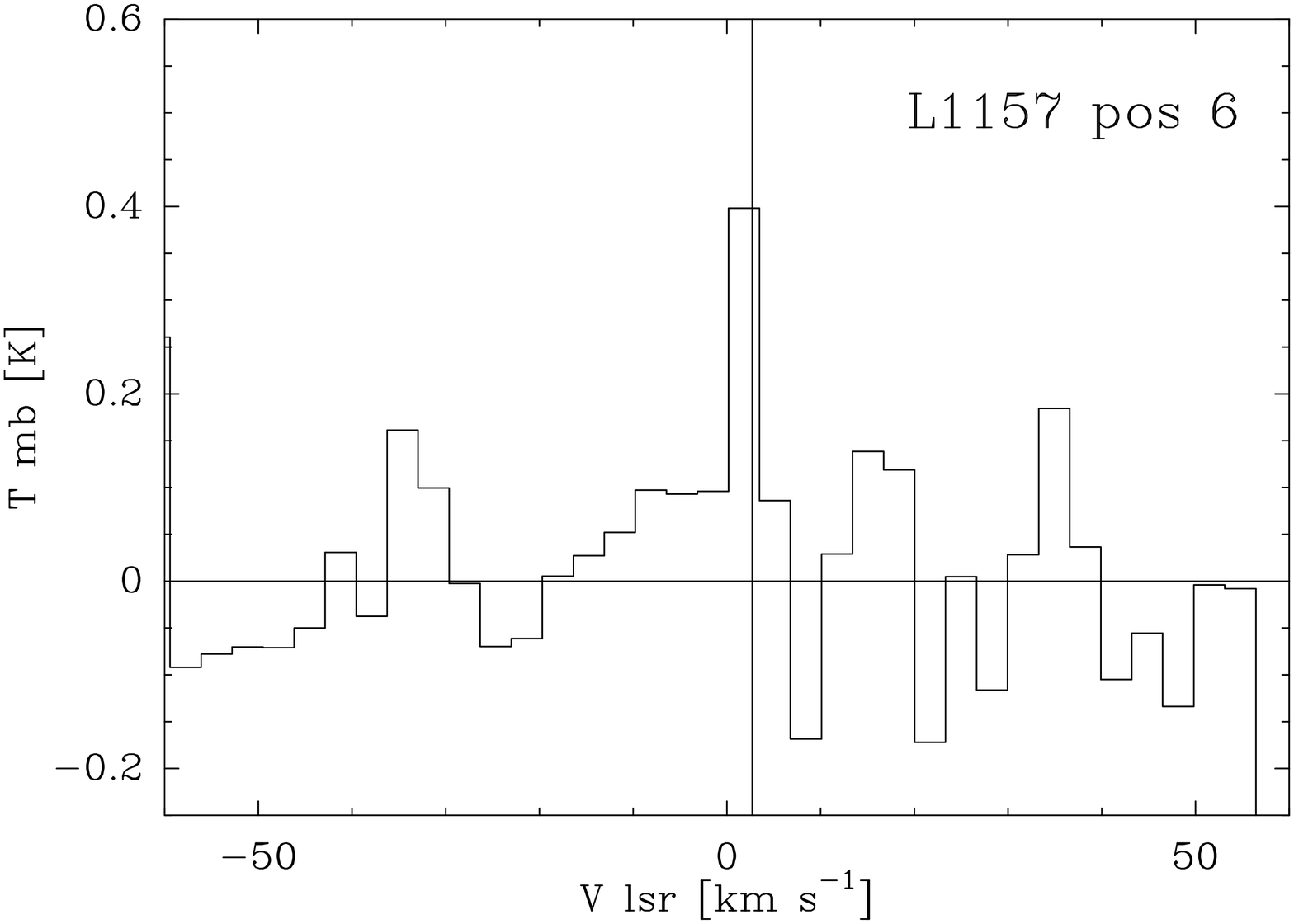}
\caption{Measurements of the CO~(12--11) line in the six positions in L1157
  marked in Fig.1. Data have been rebinned to velocity bins of
  3.3\,km\,s$^{-1}$ for these diagrams. The vertical line marks the systemic
  velcocity of $\varv _{lsr}$ = 2.7\,km\,s$^{-1}$.
\label{spectra}}
\end{figure}


\section{Observations and data reduction}

Data of the CO~(12--11) line were obtained with GREAT at various
positions in L1157. During SOFIA flights starting out of Palmdale, CA, on 
14 July 2011 positions on the central source and towards the southern 
outflow lobe were obtained, while
on the 28 and 30 September 2011 positions towards the northern outflow lobe
were observed. Fig.\,\ref{positions} shows the six targetted positions on a
H$_2$ 0--0~S(1) and CO~(2--1) map. They were selected because they are 
peaks in the SiO map of \citet{bachiller2001}.
GREAT was in its L1a/L2 configuration in upper side band.
The fast-Fourier transform spectrometers were used as backends for data
recording, and data from the XFFTS backend are shown here.

The chop amplitude was 80$''$ to the east and west of the source
position. The observations were calibrated with the chopper wheel method 
using cold and warm
loads \citep{heyminck2011}. Bandpass-averaged zenithal opacities varied from 
$\tau$ = 0.11 to 0.30, with averages of 0.21, 0.14, and 0.17 on the 
flights of 14 July, and 28 and 30, September, respectively.

The CLASS package\footnote{see http://www.iram.fr/IRAMFR/GILDAS} was used for
the standard data reduction, which  
included flagging of bad channels and atmospheric features, and a third order 
baseline subtraction. Data in the single positions were then averaged using a
sigma weighting scheme, and calibrated for a forward efficiency of 0.95 and a
main beam efficiency of 0.54. 


\begin{table*}
\caption{Measured positions and fluxes.} 
\label{measurements}
\begin{tabular}{c r r r c c c c l}
\hline
\hline
\noalign{\smallskip}
Pos & \multicolumn{2}{c}{Position offset$^{\mathrm{a}}$} &
$T_{int}$$^{\mathrm{b}}$ & rms T$_{mb}$$^{\mathrm{c}}$ & \multicolumn{2}{c}{$\int T
  d\varv$~~~~ [K km s$^{-1}$]} & flux ratio & other designations \\   
  &  $\Delta$x [$''$] & $\Delta$y [$''$] & [min]  & [K] &
CO~(12--11)$^{\mathrm{d}}$ &  CO~(2--1)$^{\mathrm{e}}$ &  CO~(12--11)/CO~(2--1)& \\
\noalign{\smallskip}
\hline
\noalign{\smallskip}
L1157-1 &  -30.2  &  140.2  & 10.0 & 0.12 & $<$1.9$^{\mathrm{f}}$~ &  64.5  &
$<$0.04~~~~~ &~~~\object{~L1157-R} \\                      
L1157-2 &  -30.2  &  114.2  & 16.0 & 0.05 &  2.5~~(0.4)  &  70.1  & 0.03~~~
(0.005) &~~~~\object{L1157-R1} \\                     
L1157-3 &  -36.5  &   54.2  &  9.3 & 0.10 &  4.5~~(0.9)  &  75.5  & 0.05~~~
(0.01)~~ &~~~~\object{L1157-R0} \\            
L1157-4 &    0.0  &    0.0  & 10.0 & 0.11 &  3.2~~(0.9)  &  42.2  & 0.07~~~
(0.02)~~ &~~~~\object{L1157-mm} \\            
L1157-5 &   22.8  &  -65.3  &  8.1 & 0.09 &  6.7~~(0.7)  &  95.1  & 0.06~~~
(0.006) &~~~~\object{L1157-B1} \\                     
L1157-6 &   35.4  & -114.8  &  7.4 & 0.16 &  2.9~~(1.3)  &  41.0  & 0.07~~~
(0.03)~~ &~~~~\object{L1157-B2} \\ 
\hline
\end{tabular}
\begin{list}{}{}
\item[] $^{\mathrm{a}}$~Position offsets are relative to the central source
  \object{L1157-mm} at R.A. 20$^h$39$^m$06\fs2 DEC 
  +68\degr 02\arcmin 15\farcs8 (J2000), 
$^{\mathrm{b}}$~total on-source integration time, 
$^{\mathrm{c}}$~measured in the range from -90\,km\,s$^{-1}$ to
  -50\,km\,s$^{-1}$ because of an atmospheric feature on the red side beyond
  +50\,km\,s$^{-1}$, 
$^{\mathrm{d}}$~this paper, 
$^{\mathrm{e}}$~from \citet{bachiller2001}, integrated from -20\,km\,s$^{-1}$ to
  +28\,km\,s$^{-1}$,
$^{\mathrm{f}}~$2-$\sigma$ upper limit.
\end{list}
\end{table*}


\section{Results}

The CO~(12--11) line was observed towards the source position
\object{L1157-mm} and five 
positions in the low-mass \object{L1157} outflow (Fig.\,\ref{positions}). 
The measured line fluxes at the various positions, integrated from 
$\varv _{lsr}$ = -20\,km\,s$^{-1}$ to +28\,km\,s$^{-1}$ as in
\citet{bachiller2001}, are given in Table\,\ref{measurements}.
CO~(12--11) was detected in the inner four positions at a signal-to-noise
ratio between 3.7 and 9.5. It was not detected in the northern-most position 1
and only marginally, at a signal-to-noise ratio of 2.3, at the southern-most
position 6 (Fig.\,\ref{spectra}). 
Although our data do not have at high signal-to-noise ratio, we find that
at positions 2 and 3 our resolved line profiles are resembling those of
\citet{bachiller2001} in CO~(2--1), with a peak near the systemic velocity at
about 5\,km\,s$^{-1}$. A second peak around 10 -- 20\,km\,s$^{-1}$, 
broader but weaker in the Bachiller et al. data, is seen in our data in 
these positions as well. A similar, but narrower peak is also present in
position 5, where a secondary component is not seen in the low-J CO spectra
(e.g. \citealt{lefloch2010}). 
e obtained the clearest detection towards position 5. In this position,
which corresponds to the bright shocked spot \object{L1157-B1}, the emission
shows a 
peak close to the systemic velocity of 2.7\,km\,s$^{-1}$, and exhibits a blue
wing that we can trace to about -20\,km\,s$^{-1}$. 

In their analysis of low-$J$ rotational CO emission in \object{L1157-B1}, 
\citet{lefloch2010} detected two velocity components in the lines: a fainter
high velocity component (HVC) blueward of $\varv _{lsr}$ =
-7.5\,km\,s$^{-1}$,  
and a brighter low velocity component (LVC) redward of $\varv _{lsr}$ =
-7.5\,km\,s$^{-1}$. Our new observations in CO~(12--11) are therefore 
mainly tracing the LVC.


\section{Discussion}

It is interesting to use the new data to estimate the physical conditions 
of the emitting gas along the \object{L1157} outflow, to infer from them 
the origin of the 
emission, and to see if these conditions are changing along the flow.

To infer the physical conditions giving rise to the CO~(12--11) 
emission observed with GREAT, we compared it with the observations 
of \citet{bachiller2001}, who mapped the entire \object{L1157} outflow in
CO~(2--1). 
Their CO~(2--1) map was obtained at the IRAM 30-m antenna with a beamsize
of 11$''$. For the purpose of our comparison we convolved this map to a 
beamsize of 21\farcs3 -- corresponding to that of GREAT for the 
CO~(12--11) line -- and measured the emission in the positions also observed  
with GREAT, integrated over a velocity range from -20\,km\,s$^{-1}$ to 
+28\,km\,s$^{-1}$, as in \citet{bachiller2001}. 
These values are reported in Table\,\ref{measurements}, col.\,7.
We integrated the emission of the CO~(12--11) line over the same velocity 
range (Table\,\ref{measurements}, col.\,6, with errors in brackets).
The corresponding line ratio of CO~(12--11)/CO~(2--1) is also reported in
Table\,\ref{measurements}. Its associated error is dominated by the
rms noise of our CO~(12--11) observations.

The first view shows that the CO~(12--11)/CO~(2--1) line ratio is fairly
constant from the source to the innermost northern knot L1157-R0 and along the
knots in the southern outflow lobe, with an average value of about 
0.06 $\pm$ 0.01. The
ratio may be decreasing for the two northern-most knots, but this difference
is not very significant, given the associated errors, and because for
the northern knot L1157-R we can only give a 3-$\sigma$ upper limit, because
of the non-detection of CO~(12--11). 

It is interesting to note that the CO line ratio of 0.07 observed for
\object{L1157-mm}, that is ``on-source'', is not significantly different from
that of the 
outflow positions. This suggests that the on-source CO emission is also
dominated by shock emission, and not by emission from the envelope of the
embedded source. We will investigate this in more detail below. 
  
To derive information about the physical conditions of the CO
emitting gas in the observed positions along the \object{L1157} outflow, we used
the radiative transfer model RADEX (\citealt{vandertak2007}) to compute 
the expected CO~(12--11)/CO~(2--1) ratio to confront it with our observations.
We adopted a plane-parallel large velocity gradient (LVG) approximation
and the collisional coefficients of \citet{yang2010}, and -- given the similar
flux ratios everywhere -- are assuming that the emission regions of both lines
must be very similar.

With a single line ratio at hand, it is obvious that the range of free
parameters has to be restricted with additional information to break
the otherwise existing strong degeneracies between gas density and
temperature. To this end, we are considering a column density of the CO gas 
that is compatible with the SPITZER observations of pure rotational
lines of molecular hydrogen of the same regions as we observed here. 
\citet{nisini2010b} estimated that the total column density of H$_2$
associated with the shocked spots is of the order of 10$^{20}$\,cm$^{-2}$, 
with variations of only a factor of a few in the various regions. 
Assuming a CO abundance of 10$^{-4}$, this translates into a column density of 
about 10$^{16}$\,cm$^{-2}$ for the CO gas. We also assumed a value for the 
velocity dispersion $\Delta \varv$ of about 10\,km\,s$^{-1}$. 
This is what we see in our spectrum at position 5 as the width of the line 
thanks to the high spectral resolution of GREAT.


\begin{figure}
\centering 
\includegraphics[width=10.5cm]{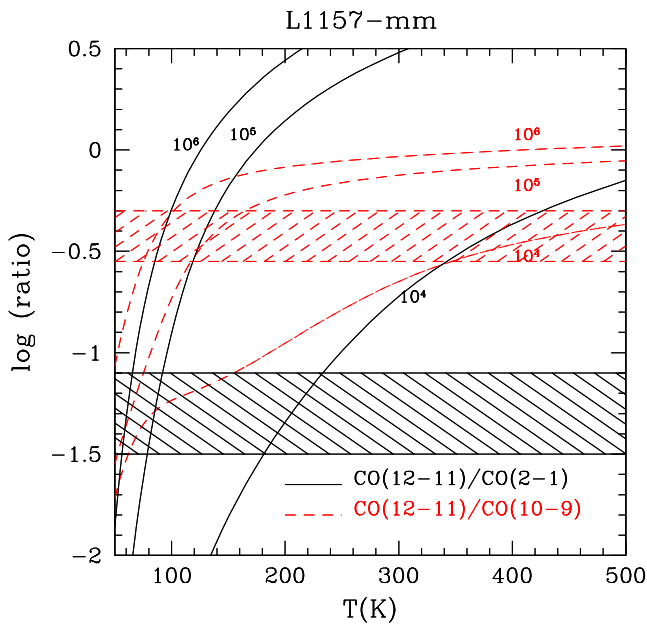}
\caption{Ratios of the integrated fluxes in CO~(12--11)/CO~(2--1)
versus gas temperature for a CO gas column density of
  10$^{16}$\,cm$^{-2}$ at the various observed positions in the \object{L1157}
  outflow 
  (in black). For the observed position of the central source \object{L1157-mm}
  the ratios for CO~(12--11)/CO~(10--9) are given as well (in red).
\label{ratio_source}}
\end{figure}


In Fig.\,\ref{ratio_source}, we show (in black) the predicted 
CO~(12--11)/CO~(2--1) line
ratio as a function of the kinetic temperature and for total H$_2$ densities of 
10$^4$, 10$^5$, and 10$^6$\,cm$^{-3}$. Because the line ratios derived from our
CO~(12--11) observations and those from the map of \citet{bachiller2001}, that
are listed in Table\,\ref{measurements} are so similar for the various
positions along the flow, they are shown as a black stripe here for clarity. 

Assuming an H$_2$ density of 10$^4$ to 10$^6$\,cm$^{-3}$, we can constrain the 
kinetic temperature of the observed CO gas to about 60 to 200\,K. In their 
analysis of the single, brightest position \object{L1157-B1},
\citet{lefloch2010} estimated a gas density of $n$(H$_2$) $\sim$
3\,$\times$\,10$^5$\,cm$^{-3}$ and a temperature of $T \sim$ 100\,K for the
gas in the low-velocity component. Comparing the estimate from our newly
observed line ratio with theirs, we find both results to be consistent. 
This confirms that we are mostly tracing the LVC component along the whole
flow, that is intermediate between the cold gas, which gets entrained into 
the flow, and the hot, shocked gas. 

We already mentioned the surprising fact that the CO~(12--11)/CO~(2--1) line
ratio at the position of the central source \object{L1157-mm} is similar to
the ratios we are finding along the flow. 
An additional constraint for the CO gas at this central position can be 
provided by an observation of the CO~(10--9) line obtained with the 
HIFI spectrometer aboard HERSCHEL (Yildiz et al., in prep.). This observation 
was obtained with a beamsize of 20$''$, which is comparable to the beamsize of 
GREAT for the CO~(12--11) line. It gives a flux integrated over the same 
velocity range as our observations of 8.4 $\pm$ 0.4\,K\,km\,s$^{-1}$. 
This then implies a CO~(12--11)/CO~(10--9) line ratio of 0.4 $\pm$ 0.1.
In Fig.\,\ref{ratio_source}, we show the model calculations with RADEX for the
line ratio of CO~(12--11)/CO~(10--9) (in red) together with that of 
CO~(12--11)/CO~(2--1) (in black). The assumed column density is again 
10$^{16}$\,cm$^{-2}$, and gas densities of 10$^4$, 10$^5$, and 10$^6$\,cm$^{-3}$ are
plotted as lines. The observed line ratios and their associated errors are
shown as black solid-lined and red dashed bands. 

Fig.\,\ref{ratio_source} shows that a low-density solution would not be able
to simultaneously account for the two line ratios: the CO~(12--11)/CO~(2--1)
ratio would imply a CO gas temperature of 200 to 240\,K, while the
CO~(12--11)/CO~(10--9) would require gas in excess of 350\,K. Assuming that both
line ratios are tracing the same bulk mass of gas, densities of between
10$^5$ and 10$^6$\,cm$^{-3}$ and gas temperatures of about 60 to 100\,K are
required. 

This analysis with the additional use of the CO~(12--11)/CO~(10--9) line ratio
thus confirms and strenghtens the gas parameters that we found from the single
line ratio along most of the \object{L1157} outflow, and the estimates of
\citet{lefloch2010} for position \object{L1157-B1} in the flow.
It supports our earlier finding that the CO emission in the on-source position
is also dominated by shocked gas, similar to what we see along the outflow, and
not by cold emission from an envelope.

\citet{hirano2001} analysed spectra of the CO~(6--5), CO~(4--3), and
CO~(3--2) lines in three different shock positions in L1157. They found that
the emission mostly comes from gas at about $n \sim$ 10$^4$\,cm$^{-3}$ and 
$T \sim$ 50 -- 150\,K. From our SOFIA observations we derive a similar
temperature range. For the central position, however, a higher density of the
order of 10$^5$ to 10$^6$\,cm$^{-3}$ is indicated. 

\citet{benedettini2012} have analysed the velocity-integrated CO emission from
J = 14--13 to J = 22--21 observed in position B1 (our position 5 ) with
HERSCHEL. They derived that these lines are compatible with gas at 
200 $< T<$ 800\,K and n $\ge$ 10$^5$ cm$^{-3}$.
Clearly, CO emission at higher J is dominated by a gas warmer than what
we are inferring here, testifying that the emission of the series of lines of 
the CO ladder is able to trace the temperature stratification that is 
present across the shocked gas in the outflow.


\section{Conclusions}

We presented observations of the CO~(12--11) line at various positions along 
the low-mass \object{L1157} bipolar outflow obtained with GREAT aboard SOFIA.

Comparing these new CO~(12--11) data with CO~(2--1), we found that the line
ratios are nearly constant along the flow, even at the position of the outflow
source. An estimate of the physical parameters of the emitting gas shows that
we are most likely tracing matter with a total gas density of about 10$^5$ to
10$^6$\,cm$^{-3}$ and at a temperature between 60 to 100\,K.

This indicates that we are mostly tracing a low-velocity gas component in the
outflow, which is intermediate between the cold gas that gets entrained into
the outflow and is seen in low-$J$ rotational CO lines, and the hot gas in the
shocks in 
the outflow that was seen in very high-$J$ rotational CO lines with ISO.


\begin{acknowledgements}
We thank Rafael Bachiller for making his CO~(2--1) map of \object{L1157}
available to us. 
We thank Umut Yildiz for making the information on the CO~(10--9) integrated
emission available to us prior to publication.
Based [in part] on observations made with the NASA/DLR Stratospheric
Observatory for Infrared Astronomy. SOFIA Science Mission Operations are
conducted jointly by the Universities Space Research Association, Inc., under
NASA 
contract NAS2-97001, and the Deutsches SOFIA Institut und DLR contract 50 OK
0901. JE is grateful for a travel grant from the German Aerospace Center DLR,
which allowed him to be onboard SOFIA for the first flight, on which data for 
this programme were obtained.

\end{acknowledgements}




\begin{thebibliography}{}

\bibitem[Bachiller~et~al.(2001)]{bachiller2001} Bachiller, R., 
   P\'erez Guti\'errez, M., Kumar, M.S.N., Tafalla, M. 2001,
   A\&A 372, 899

\bibitem[Benedettini~et~al.(2012)]{benedettini2012} Benedettini, M.,
   Busquet G., Lefloch B. et al. 2012,
   A\&A, in press

\bibitem[Caratti o Garatti~et~al.(2006)]{caratti2006} Caratti o Garatti, A., 
   Giannini, T., Nisini, B., Lorenzetti, D. 2006,
   A\&A 449, 1077

\bibitem[Codella~et~al.(2010)]{codella2010} Codella, C., Lefloch, B., 
   Ceccarelli, C., Cernicharo, J., Caux, E., Lorenzani, A., Viti, S. et al. 2010,
   A\&A, 518, L112

\bibitem[Davis \& Eisl\"offel(1995)]{daviseisloeffel1995} Davis, C.J., 
   Eisl\"offel, J. 1995, 
   A\&A 300, 851

\bibitem[Giannini~et~al.(2001)]{giannini2001}  Giannini, T., Nisini, B., 
   Lorenzetti, D. 2001,
   ApJ 555, 40

\bibitem[Heyminck~et~al.(2012)]{heyminck2011} Heyminck, S., Graf, U.U., 
   G\"usten, R., Stutzki, J., H\"ubers, H.W., Hartogh, P. 2011,
   A\&A, this issue

\bibitem[Hirano \& Taniguchi(2001)]{hirano2001} Hirano, N., Taniguchi, Y. 2001,
   ApJ 550, L219

\bibitem[Lefloch~et~al.(2010)]{lefloch2010} Lefloch, B., Cabrit, S., Codella, C., 
   Melnick, G., Cernicharo, J., Caux, E., Benedettini, M., et al. 2010, 
   A\&A, 518, L113

\bibitem[Nisini~et~al.(2010a)]{nisini2010a}  Nisini, B., Benedettini, M., Codella, C.,
   Giannini, T., Liseau, R., Neufeld, D.A., Tafalla, M. et al. 2010,
   A\&A 518, L120

\bibitem[Nisini~et~al.(2010b)]{nisini2010b}  Nisini, B., Giannini, T., Neufeld, D.A.,
   Yuan, Y., Antoniucci, S., Bergin, E.A., Melnick, G.J. 2010b, 
   ApJ 724, 69

\bibitem[van der Tak~et~al.(2007)]{vandertak2007} van der Tak, F.F.S., Black,
   J.H., Sch\"oier, F.L., Jansen, D.J., van Dishoeck, E.F. 2007, 
   A\&A 468, 627

\bibitem[Yang~et~al.(2010)]{yang2010} Yang, B., Stancil, P.C., Balakrishnan, N., 
   Forrey, R.C. 2010, 
   ApJ 718, 106

\bibitem[Young~et~al.(2012)]{young2011} Young, E.T, Becklin, E.E., De Buizer, J.M.,
   Herter, T.L., G\"usten, R., Dunham, E.W., Sankrit, R., et al. 2011,
   ApJ Letters, in press


\end{thebibliography}
\end{document}